\documentclass[amsmath,amssymb,aps,pre,twocolumn,groupedaddress]{revtex4-2}
\usepackage[T1]{fontenc}
\usepackage{graphicx}
\usepackage{color}
\usepackage{bm}
\usepackage{amsmath}
\usepackage{slashed}
\usepackage{mathrsfs}
\usepackage[normalem]{ulem}
\usepackage{comment}

\usepackage{dcolumn}% Align table columns on decimal point
\usepackage{hyperref}
\hypersetup{colorlinks=true, citecolor=blue, urlcolor=blue, linkcolor=blue}
\usepackage{epstopdf}

\begin{document}

\title{Decomposition of metric tensor in thermodynamic geometry in terms of relaxation timescales}

\author{Zhen Li}
\email{li-zhen@g.ecc.u-tokyo.ac.jp}
\affiliation{Department of Complexity Science and Engineering, Graduate School of Frontier Sciences, The University of Tokyo, Kashiwa 277-8561, Japan.}
\author{Yuki Izumida}
\email{izumida@k.u-tokyo.ac.jp}
\affiliation{Department of Complexity Science and Engineering, Graduate School of Frontier Sciences, The University of Tokyo, Kashiwa 277-8561, Japan.}

\begin{abstract}
    Geometrical methods are extensively applied to thermodynamics including stochastic thermodynamics.
    In the case of slow-driving linear response regime, a geometrical framework, known as thermodynamic geometry, is established.
    The key of this framework is the thermodynamic length characterized by a metric tensor defined on the space of controlling variables.
    %It is shown that the metric tensor can be decomposed into a kind of product between a timescale matrix and the Fisher information matrix.
    As the metric tensor is given in terms of the equilibrium time-correlation functions of the thermodynamic forces, it contains the information of timescales, which may be useful for analyzing the performance of heat engines.
    In this paper, we show that the metric tensor for underdamped Langevin dynamics can be decomposed in terms of the relaxation times of a system itself, which govern the timescales of the equilibrium time-correlation functions of the thermodynamic forces.
    As an application of the decomposition of the metric tensor, we demonstrate that it is possible to achieve Carnot efficiency at finite power by taking the vanishing limit of relaxation times without breaking trade-off relations between efficiency and power of heat engines in terms of thermodynamic geometry.
\end{abstract}
\maketitle

\section{Introduction}\label{sec_intro}

Geometrical methods are widely introduced into thermodynamics~\cite{gibbs1928collected,weinhold1975metric,ruppeiner1979thermodynamics,hernandez1998contact,salamon1983thermodynamic,crooks2007measuring,Blaber_2023,Deffner_2020}.
They include stochastic thermodynamics, which has developed as thermodynamics for fluctuating systems~\cite{sekimoto2010stochastic,Seifert_2012,Peliti2021stochastic,shiraishi2023introduction}.
As one of the most important findings, entropy production or dissipation availability as the irreversible energy loss for finite-time thermodynamic processes is known to be bounded by geometric quantities~\cite{Blaber_2023,Deffner_2020}. 
Among others, for the slow driving linear response regime in which the system's relaxation time is much shorter than the duration of the driving process, the thermodynamic length in thermodynamic geometry serves as such a geometrical quantity, which describes distance between points in the space of control parameters~\cite{salamon1983thermodynamic,crooks2007measuring,sivak2012thermodynamic,Blaber_2023,Deffner_2020,Peliti2021stochastic}.
The thermodynamic length is characterized by a metric tensor defined by the equilibrium time-correlation functions of the thermodynamic forces~\cite{sivak2012thermodynamic}, and also works as the coefficients of the linear response relations between the control variables and their conjugate thermodynamic forces~\cite{brandner2020thermodynamic,frim2022geometric}.
The thermodynamic geometry is successfully applied to the optimization of various thermodynamic processes~\cite{zulkowski2012geometry,rotskoff2015optimal,sivak2016thermodynamic,zulkowski2014optimal,scandi2022minimally,frim2022geometric,frim2022optimal,brandner2020thermodynamic,abiuso2020optimal,miller2020,watanabe2022finite,zhang2024geometry}.

Another but tightly related geometric approach is based on optimal transport theory~\cite{villani2003}.
In this approach, the entropy production is bounded by Wasserstein distance defined in terms of the optimal cost for transporting probability distribution from an initial distribution to a final one~\cite{Aurell2011,Aurell2012,Van_Vu_2021,nakazato2021geometrical,Van_Vu2023,ito2024geometric}. See also, e.g., Refs.~\cite{proesmans2020finite,van2022finite,vu2024geometric,Sabbagh2024,Nagase2024} for other interesting applications.
Remarkably, for overdamped Langevin dynamics, it is shown that  
optimal transport geometry 
has a tight relationship with thermodynamic geometry 
even beyond the linear response regime~\cite{chennakesavalu2023unified,zhong2024beyond},
sharing identical geodesics and distances~\cite{zhong2024beyond}.

Though there are some theories connected to optimal transport that are applied in optimizations~\cite{schmiedl2007optimal, proesmans2020finite, van2022finite}, such theories in underdamped dynamics may not be as simple as in overdamped cases~\cite{sanders2024optimal,sanders2024minimum}.
Thus, the thermodynamic length in thermodynamic geometry serves as an efficient tool in optimizing thermodynamic systems including heat engines as an important example~\cite{brandner2020thermodynamic,miller2020,abiuso2020optimal,watanabe2022finite,frim2022geometric,frim2022optimal}.

The achievement of Carnot efficiency is usually unrealistic due to the infinitely long operation time to make all the processes quasi-static~\cite{curzon1975efficiency,andresen1984thermodynamics,andresen2011current}, and the infinitely long operation time obviously leads to vanishing power, making the heat engine impractical.
The amount of performance decrease we should be tolerant of was a key question in finite-time thermodynamics~\cite{andresen2011current},
and it was recently quantified as trade-off relations between efficiency and power of heat engines~\cite{shiraishi2016universial,pietzonka2018universal,dechant2018entropic,brandner2020thermodynamic,hino2021geometrical,vu2024geometric}.
In particular, in the slow driving linear response regime, the thermodynamic length with the metric tensor plays an important role in the trade-off relation~\cite{brandner2020thermodynamic}.

Meanwhile, it was pointed out recently that the vanishing limit of relaxation times of a system could lead to the compatibility of Carnot efficiency and finite power without breaking the trade-off relation~\cite{holubec2018cycling,miura2021compatibility,miura2022achieving}.
Because the metric tensor includes the relaxation timescales of the equilibrium time-correlation functions~\cite{sivak2012thermodynamic,watanabe2022finite,sawchuk2024dynamical}, it may be useful for the analysis of the performance of heat engines as the relaxation times significantly affect the efficiency and power of heat engines. Furthermore, it is expected that the compatibility of Carnot efficiency and finite power may be explained in terms of the property of the metric tensor in the slow driving linear response regime.

In this paper, we explore the detailed structure of the metric tensor in the framework of stochastic thermodynamics and apply it to the problem of compatibility of Carnot efficiency and finite power in stochastic heat engines. As the relaxation dynamics of the correlation between thermodynamic forces fundamentally reflect the dynamics of a system obeying Langevin equations, it is natural that the metric tensor is decomposed in terms of the relaxation times of the system, such as those of position and momentum.
We demonstrate this decomposition analytically for a harmonic potential and numerically for scale-invariant potentials by using Langevin dynamics.
As an application of the decomposition of the metric tensor, we show that the compatibility of Carnot efficiency and finite power is achieved in the vanishing limit of relaxation times, by analyzing the dissipated availability with the aid of the decomposition on the metric tensor. Furthermore, it will be shown that the compatibility is consistent with the trade-off relations between efficiency and power.

The organization of the rest of the paper is as follows. We first give an introduction to the metric tensor in thermodynamic geometry in \autoref{sec_metric}.
Then, we demonstrate the decomposition of the metric tensor according to different relaxation timescales in \autoref{sec_decomposition}.
As an application of the decomposition of the metric tensor, we show the compatibility of Carnot efficiency and finite power and its consistency with the trade-off relation between efficiency and power in~\autoref{sec_compatibility}.
Finally, we give concluding remarks in \autoref{sec_discussion}.

\section{Metric tensor in thermodynamic geometry}
\label{sec_metric}
Let us consider the dynamics of a Brownian particle described by the following underdamped Langevin equations for position $x$ and momentum $p$:
\begin{align}
    \dot{x} &= \frac{p}{m},\label{eq_dx}\\
    \dot{p} &= -\frac{\partial V}{\partial x} - \frac{\xi}{m}p + \zeta(t).\label{eq_dp}
\end{align}
Here, the dot denotes the time derivative.
$m$, $\xi$, and $V=V(x,\Lambda^i)$ denote the mass of the particle, friction coefficient, and a potential function with $\Lambda^i$ ($i=1,2,\cdots, M$) being $M$ time-dependent parameters as ``mechanical" control variables, respectively.
$\zeta(t)$ is the Gaussian white noise satisfying the fluctuation-dissipation relation~\cite{kadanoff2000statistical,risken1996fokker,kubo1966fluctuation}:
\begin{equation}
\left<\zeta(t)\right>=0, \quad  \left<\zeta(t)\zeta(t')\right> = 2\xi k_{\rm B} T(t) \delta(t-t'),
\end{equation}
where the bracket $\left<\cdots\right>$ denotes the ensemble average, $T(t)$ is the time-dependent temperature of the heat reservoir as a ``thermal" control variable, and $k_{\rm B}$ is Boltzmann constant.
We collectively write all control variables as $\Lambda^\mu \equiv (T,\Lambda^i)$.
The Brownian particle may be regarded as the working substance of a heat engine if temperature $T$ and parameters $\Lambda^i$ in the potential $V$ are changed periodically.

Following the framework in~\cite{sivak2012thermodynamic,watanabe2022finite,brandner2020thermodynamic}, we can derive the linear response relations between the conjugated thermodynamic forces $X_\mu$ and the changing rates of control variables $\Lambda^\mu$:
\begin{align}
    \left<\delta X_\mu\right> \equiv \left<X_\mu\right>-\mathcal{X}_\mu = -g_{\mu\nu}\dot{\Lambda}^{\nu},\label{eq_linear_response}
\end{align}
where $\mathcal{X}_{\mu} \equiv \left<X_\mu\right>_{\rm eq}$ is the quasi-static value of $X_\mu$ with $\left<\cdots \right>_{\rm eq}$ denoting the ensemble average at equilibrium, and $\delta X_\mu \equiv X_\mu - \mathcal{X}_\mu$ is the fluctuation of $X_\mu$.
The thermodynamic force $X_i$ conjugated to $\Lambda^i$ is given by
\begin{equation}
    X_i \equiv -\frac{\partial H}{\partial \Lambda^i}\label{eq_Xi_def}
\end{equation}
as a generalized force, where $H \equiv p^2/(2m) + V$ is the system's Hamiltonian.
Meanwhile, $X_T$ conjugated to $T$ is given by
\begin{equation}
    X_T \equiv -k_{\rm B} \ln \rho \label{eq_XT_def}
\end{equation}
as a stochastic entropy, where {$\rho_{\rm eq}=\rho_{\rm eq}(x,p)$ is the distribution of the system.
The linear response coefficients $g_{\mu\nu}$ are given by the equilibrium time-correlation functions of the thermodynamic forces~\cite{sivak2012thermodynamic}:
\begin{equation}
    g_{\mu\nu} \equiv \frac{1}{k_{\rm B} T}\int_{0}^{+\infty}ds \left<\delta X_\mu(s)\delta X_\nu(0)\right>_{\rm eq}.\label{eq_g_def}
\end{equation}

An operator solution for the time evolution of a function $\phi(s)$ can be given as~\cite{watanabe2022finite}:
\begin{equation}
    \phi(s) = e^{\mathcal{L}_{\rm FP}^\dagger s}\phi(0),\label{eq_phi(s)}
\end{equation}
where
\begin{equation}
    \mathcal{L}_{\rm FP}^\dagger \equiv \frac{p}{m}\frac{\partial}{\partial x} - \frac{\partial V}{\partial x}\frac{\partial}{\partial p} - \frac{\xi}{m}p\frac{\partial}{\partial p} + \xi k_{\rm B} T \frac{\partial^2}{\partial p^2},\label{eq_ad_fp}
\end{equation}
is the adjoint Fokker-Planck operator corresponding to the Langevin equations \eqref{eq_dx} and \eqref{eq_dp}~\cite{kadanoff2000statistical,risken1996fokker}.
With this operator solution, we can find another expression for $\delta X_T(s)$:
\begin{align}
    \delta X_T(s) &= e^{\mathcal{L}_{\rm FP}^\dagger s}\delta X_T (0)\notag\\
    &= e^{\mathcal{L}_{\rm FP}^\dagger s}\left(-k_{\rm B} \ln\rho_{\rm eq} +k_{\rm B}\left<\ln\rho_{\rm eq}\right>_{\rm eq}\right)\notag\\
    &= e^{\mathcal{L}_{\rm FP}^\dagger s}\left(\frac{H(0)}{T} - \frac{\mathcal{F}}{T} - \mathcal{S}\right)\notag\\
    &= e^{\mathcal{L}_{\rm FP}^\dagger s}\left(\frac{H(0)}{T} - \frac{\mathcal{U}}{T}\right)\notag\\
    &= \frac{H(s)}{T} - \frac{\mathcal{U}}{T},\label{eq_XT_re}
\end{align}
if the system is initialized with the equilibrium distribution:
\begin{equation}
    \rho_{\rm eq} \equiv \exp\left\{-\frac{1}{k_{\rm B} T}\left(H-\mathcal{F}\right)\right\},\label{eq_distri}
\end{equation}
where $\mathcal{F}\equiv \mathcal{U} - T\mathcal{S}$ is the Helmholtz free energy with $\mathcal{U}\equiv \left<H\right>_{\rm eq}$ and $\mathcal{S}\equiv -k_{\rm B} \left<\ln\rho_{\rm eq}\right>_{\rm eq}$ being the equilibrium internal energy and the equilibrium entropy, respectively.

It is also worth noting that the fluctuations of thermodynamic forces are given by
\begin{equation}
    \delta X_{\mu}(0) = k_{\rm B} T \frac{\partial \ln \rho_{\rm eq}}{\partial \Lambda^{\mu}},\label{eq_dX_rho}
\end{equation}
if we initialize the system as the equilibrium state~\cite{watanabe2022finite}.

In the slow-driving linear response regime, the dissipated availability $\mathcal{A}$ for a process starting at time $t=t_i$ and ending at time $t=t_f$ is given in a geometrical way in terms of $g_{\mu\nu}$:
\begin{equation}
    \mathcal{A} \equiv \int_{t_i}^{t_f} g_{\mu\nu}\dot{\Lambda}^{\mu}\dot{\Lambda}^{\nu}dt.\label{eq_diss}
\end{equation}
As a result of the second law of thermodynamics, we have $\mathcal{A}\geq 0$, requiring the positive semi-definiteness of $g_{\mu\nu}$.

There is a relation between dissipated availability $\mathcal{A}$ and entropy generation rate $\dot\sigma$:
\begin{equation}
    \mathcal{A}=\int_{t_i}^{t_f} T\dot \sigma dt,\label{eq_entropy}
\end{equation}
where $\dot \sigma \equiv \dot S - J / T$
with $S = \left<X_T\right>$ and $J$ being the entropy and the heat flux, respectively.
To derive Eq.~\eqref{eq_entropy}, we start from the first law of thermodynamics:
\begin{equation}
    \frac{d}{dt}\left<H\right> = J - \left<X_i\right> \dot{\Lambda}^i, \label{eq_1st_law}
\end{equation}
where the changing rate of internal energy $d\left<H\right>/dt$ is equated with the heat flux $J$ minus the instantaneous power $\left<X_i\right> \dot{\Lambda}^i$.
Based on Eq.~\eqref{eq_1st_law}, we have~\cite{brandner2020thermodynamic}
\begin{equation}
    -\left<X_\mu\right> \dot{\Lambda}^\mu = T\dot{\sigma} + \frac{d}{dt}\left(\left<H\right> - TS\right).\label{eq_brandner}
\end{equation}
Combined with Eq.~\eqref{eq_XT_re}, Eq.~\eqref{eq_brandner} becomes
\begin{equation}
    -\mathcal{X}_\mu \dot{\Lambda}^\mu +g_{\mu\nu}\dot{\Lambda}^\mu\dot{\Lambda}^\nu = T\dot{\sigma} + \frac{d}{dt}\left(\mathcal{U}-T\mathcal{S}\right),
\end{equation}
by applying the linear response relation~\eqref{eq_linear_response}.
Due to the fact that $d\mathcal{F} = d(\mathcal{U} - T\mathcal{S}) = -\mathcal{X}_\mu d\Lambda^\mu$, we finally have
\begin{equation}
    T\dot{\sigma} = g_{\mu\nu}\dot{\Lambda}^\mu\dot{\Lambda}^\nu,
\end{equation}
and Eq.~\eqref{eq_entropy} combined with the definition of dissipated availability in Eq.~\eqref{eq_diss}.

By applying the Cauchy-Schwarz inequality to Eq.~(\ref{eq_diss}), we can find the lower bound of $\mathcal{A}$ for a given process as
\begin{align}
    \mathcal{A} &\geq \frac{\mathcal{L}^2}{{t_i}-{t_f}},\label{eq_cauchy}
\end{align}
where 
\begin{align}    
    \mathcal{L} &\equiv \int_{t_i}^{t_f} \sqrt{g_{\mu\nu}\dot{\Lambda}^{\mu}\dot{\Lambda}^{\nu}}dt \label{eq_length}
\end{align}
is the thermodynamic length of the trajectory corresponding to the process in $\Lambda^\mu$ space with $g_{\mu\nu}$ serving as a metric tensor.

For later use, we also introduce the overdamped Langevin equation as
\begin{equation}
    \dot{x} = -\frac{1}{\xi}\frac{\partial V}{\partial x} + \frac{1}{\xi}\zeta(t),\label{eq_dx_od}
\end{equation}
which can be derived from the underdamped Langevin equations (\ref{eq_dx}) and (\ref{eq_dp}) by assuming timescale separation $\tau_p/\tau_x\ll 1$
with $\tau_p$ and $\tau_x$ being the relaxation times of momentum and position of the particle, respectively.
While the definition of $g_{\mu\nu}$ in this overdamped description is essentially the same as the underdamped one in Eq.~(\ref{eq_g_def}), the Hamiltonian is replaced by $H = V$ and the equilibrium distribution $\rho_{\rm eq}$ in Eq.~(\ref{eq_distri}) accordingly.
Moreover, the adjoint Fokker-Planck operator in Eq.~\eqref{eq_ad_fp} should also be replaced as
\begin{equation}
    \mathcal{L}_{\rm FP}^\dagger\to\mathcal{G}_{\rm FP}^\dagger \equiv -\frac{1}{\xi}\frac{\partial V}{\partial x}\frac{\partial}{\partial x} + \frac{k_{\rm B} T}{\xi}\frac{\partial^2}{\partial x^2}.\label{eq_ad_fp_od}
\end{equation}

\section{Decomposition of metric tensor}
\label{sec_decomposition}
The metric tensor~\eqref{eq_g_def} can be decomposed into the Hadamard product of the relaxation time matrix $\tau_{\mu\nu}$ and the Fisher information matrix $I_{\mu\nu}$~\cite{sivak2012thermodynamic,watanabe2022finite}:
\begin{align}
    g_{\mu\nu} &= k_{\rm B}T\tau_{\mu\nu}\odot I_{\mu\nu},\label{eq_handamard}\\
    I_{\mu\nu} &\equiv \left<\left(\frac{\partial \ln\rho_{\rm eq}}{\partial \Lambda^\mu}\right)\left(\frac{\partial \ln\rho_{\rm eq}}{\partial \Lambda^\nu}\right)\right>_{\rm eq}.\label{eq_Fisher}
\end{align}
The Hadamard product results in a matrix where each element is the product of the corresponding element in the original matrices, 
and there is no dependence on timescales for $I_{\mu\nu}$.
Here, each element of $\tau_{\mu \nu}$ gives the correlation time between thermodynamic forces $\delta X_\mu$ and $\delta X_\nu$~\cite{sivak2012thermodynamic}.
This decomposition naturally comes from the definition~\eqref{eq_g_def} with $ \delta X_\mu(s) = e^{\mathcal{L}_{\rm FP}^\dagger s}\delta X_\mu(0)$, 
where the relaxation time of the correlation appears from the integration of the time-correlation function,
and the time-correlation functions at $s=0$ take the form of the Fisher information when considering Eq.~\eqref{eq_dX_rho}.
As multiple timescales, the relaxation times of momentum $\tau_p$ and position $\tau_x$, appear in the Langevin system~\eqref{eq_dx} and \eqref{eq_dp}, 
it is natural to consider that each component of $\tau_{\mu \nu}$ may be decomposed in terms of $\tau_p$  and $\tau_x$, which will be demonstrated in the following subsections.

\subsection{Case of harmonic potential}
Due to the linearity of Langevin equations~\eqref{eq_dx} and~\eqref{eq_dp}~\cite{frim2022optimal,zulkowski2012geometry}, and as the distribution remains Gaussian type when initialized as the equilibrium one~\cite{frim2022optimal}, it is relatively easy to calculate the metric tensor $g_{\mu\nu}$ for a harmonic potential $V(x, k)=kx^2/2$.
The position and momentum relaxation times $\tau_x$ and $\tau_p$ of this case read
\begin{align}
&\tau_x\equiv\frac{\xi}{k},\label{eq.tau_x}\\
&\tau_p\equiv\frac{m}{\xi}.\label{eq.tau_p}
\end{align}
For the control variables $\Lambda^\mu = (T,k)$, $g_{\mu\nu}$ is given by~\cite{frim2022optimal,frim2022geometric,zulkowski2012geometry}:
%\begin{widetext}
    \begin{align}
        g_{\mu\nu} = 
        \begin{pmatrix}
            g_{TT} & g_{Tk}\\
            \\
            g_{kT} & g_{kk}
        \end{pmatrix}_{\mu\nu}
        =
        \begin{pmatrix}
            \frac{mk_{\rm B}}{\xi T} + \frac{\xi k_{\rm B}}{4kT} & -\frac{mk_{\rm B}}{2\xi k}-\frac{\xi k_{\rm B}}{4k^2}\\
            \\
            -\frac{mk_{\rm B}}{2\xi k}-\frac{\xi k_{\rm B}}{4k^2} & \frac{mk_{\rm B} T}{4\xi k^2}+\frac{\xi k_{\rm B} T}{4k^3}
        \end{pmatrix}_{\mu\nu}.\label{eq_g_harmonic}
    \end{align}
%\end{widetext}
Because the Fisher information in Eq.~(\ref{eq_Fisher}) is given by
\begin{equation}
    I_{\mu\nu}=
    \begin{pmatrix}
        \frac{1}{T^2} & -\frac{1}{2Tk}\\
        \\
        -\frac{1}{2Tk} & \frac{1}{2k^2}
    \end{pmatrix},
\end{equation}
the relaxation time matrix $\tau_{\mu\nu}$ is written as
\begin{equation}
    \tau_{\mu\nu} =
    \begin{pmatrix}
        \frac{m}{\xi} + \frac{\xi}{4k} & \frac{m}{\xi} + \frac{\xi}{2k}\\
        \\
        \frac{m}{\xi} + \frac{\xi}{2k} & \frac{m}{2\xi} + \frac{\xi}{2k}
    \end{pmatrix}
    =
    \begin{pmatrix}
        \tau_p +\frac{1}{4}\tau_x & \tau_p +\frac{1}{2}\tau_x\\
        \\
        \tau_p +\frac{1}{2}\tau_x & \frac{1}{2}\tau_p +\frac{1}{2}\tau_x
    \end{pmatrix}
\end{equation}
in terms of the linear combination of $\tau_x$ and $\tau_p$.
The metric tensor $g_{\mu\nu}$ in Eq.~(\ref{eq_g_harmonic}) is positive definite, which can be checked from the positivity of its principal submatrices ($g_{TT}>0$ and ${\rm det}(g_{\mu\nu})=mk_{\rm B}^2 / (16k^3)>0$).
It is easy to find that the positive definite $g_{\mu\nu}$ is decomposed into two positive semi-definite parts:
\begin{align}
g_{\mu\nu} = g_{\mu\nu}^p + g_{\mu\nu}^x,\label{eq_g_harmonic_decompose}
\end{align}
where 
\begin{equation}
    g_{\mu\nu}^p=\tau_p
    \begin{pmatrix}
        \frac{k_{\rm B}}{ T} & -\frac{k_{\rm B}}{ 2k}\\
        \\
        -\frac{k_{\rm B}}{2 k} & \frac{k_{\rm B} T}{4k^2}
    \end{pmatrix}_{\mu\nu}\label{eq_g_harmonic_p}
\end{equation}
is proportional to the momentum relaxation time $\tau_p$, and
\begin{equation}
    g_{\mu\nu}^x=\tau_x
    \begin{pmatrix}
        \frac{k_{\rm B}}{4T} & -\frac{k_{\rm B}}{4k}\\
        \\
        -\frac{k_{\rm B}}{4k} & \frac{k_{\rm B} T}{4k^2}
    \end{pmatrix}_{\mu\nu}\label{eq_g_harmonic_x}
\end{equation}
is proportional to the position relaxation time $\tau_x$.

Here, we find that $g_{\mu\nu}^p$ and $g_{\mu\nu}^x$, respectively, are degenerate and have zero-eigenvalue.
The eigenvector $(T, 2k)$ subject to the zero-eigenvalue of $g_{\mu\nu}^p$ defines the direction of the underdamped adiabatic process 
of the Langevin system~\eqref{eq_dx} and \eqref{eq_dp}~\cite{frim2022geometric}; we obtain the adiabatic curve $T^2/k = {\rm const.}$ 
by solving $\dot{T}/\dot{k}=dT/dk = T/(2k)$. 
Meanwhile, the eigenvector $(T,k)$ for $g_{\mu\nu}^x$ defines the direction of the adiabatic process of the overdamped Langevin system~\eqref{eq_dx_od}~\cite{watanabe2022finite};
we obtain the adiabatic curve $T/k = {\rm const.}$ by solving $\dot{T}/\dot{k}=dT/dk = T/k$.
Here, the term ``adiabatic'' means $d\mathcal{S} = 0$ for the quasi-static entropy $\mathcal{S}$.

Furthermore, $g_{\mu\nu}^x$ in Eq.~(\ref{eq_g_harmonic_x}) is equal to the metric tensor $g_{\mu\nu}^{\rm od}$ for the overdamped dynamics~\eqref{eq_dx_od} for the harmonic potential~\cite{watanabe2022finite}.
This is consistent with the fact that Eq.~\eqref{eq_dx_od} is obtained in the overdamped limit $\tau_p/\tau_x \ll 1$ and from the explicit form of Eqs.~(\ref{eq_g_harmonic_decompose})--(\ref{eq_g_harmonic_x}).
In fact, two properties, $g_{\mu\nu}\to g_{\mu\nu}^{\rm od}$ in the sufficiently large $\xi$ limit and $g_{\mu\nu}^{\rm od}\propto \tau_x$, can be shown apart from the harmonic potential case, see Appendix~\ref{appendix_overdamped} and Appendix~\ref{appendix_overdamped_tau}, respectively.

\subsection{Case of scale-invariant potentials}
Inspired by the case of the harmonic potential, we conjecture that the metric tensor for other potential functions is also decomposed in the same form
in terms of the relaxation times $\tau_p$ and $\tau_x$ as
\begin{align}
    g_{\mu\nu} &= g_{\mu\nu}^p + g_{\mu\nu}^x,\label{eq_dem1}
\end{align}
where
\begin{align}    
    g_{\mu\nu}^p &\propto \tau_p,\\
    g_{\mu\nu}^x &\propto \tau_x.\label{eq_dem3}
\end{align}
However, due to the nonlinear term $\partial V/\partial x$ in Eq.~\eqref{eq_dp}, it is hard to directly calculate the metric tensor and decompose it.
Here, we provide simulation results to show numerical evidence to support our conjecture.

We consider the metric tensor for the following scale-invariant potentials: 
\begin{align}
V=\frac{kx^{2n}}{2n} \ (n\ge 1), 
\end{align}
and control variables $\Lambda^\mu = (T,k)$ as examples.
By dimensional analysis, we can identify the position relaxation timescale $\tau_x$ as
\begin{equation}
    \tau_x=\xi k^{-\frac{1}{n}}(k_{\rm B}T)^{-\frac{n-1}{n}},\label{eq_tau_x}
\end{equation}
while $\tau_p$ is the same as the case of the harmonic potential in Eq.~(\ref{eq.tau_p}).
It should be noted that the mass $m$ should not appear in the position relaxation timescale, which is related to the mass-independent overdamped Langevin dynamics~\eqref{eq_dx_od}.

We nondimensionalize the equations for the numerical simulations.
$T$ and $k$ can be nondimensionalized by setting a unit $T$ as $T_e$ and a unit $k$ as $k_e$:
\begin{align}
    \tilde{T} &\equiv T_e^{-1}T,\\
    \tilde{k} &\equiv k_e^{-1}k.
\end{align}
Then, we nondimensionalize other quantities as
\begin{align}
\tilde{x} &\equiv (k_{\rm B} T_e)^{-\frac{1}{2n}}k_e^{\frac{1}{2n}}x,\\
\tilde{p} &\equiv (mk_{B} T_e)^{-\frac{1}{2}} p,\\
\tilde{t} &\equiv m^{-\frac{1}{2}} k_e^{\frac{1}{2n}}(k_{\rm B} T_e)^{\frac{n-1}{2n}}t,\\
\tilde{\xi} &\equiv m^{-\frac{1}{2}} k_e^{-\frac{1}{2n}}(k_{\rm B} T_e)^{-\frac{n-1}{2n}} \xi,\\
\tilde{V} &\equiv (k_{\rm B} T_e)^{-1} V\\
\tilde{\zeta}(\tilde{t}) &\equiv k_e^{-\frac{1}{2n}}(k_{\rm B} T_e)^{-\frac{2n-1}{2n}}\zeta(t),
\end{align}
where $\left<\tilde \zeta(t) \right>=0$ and $\left<\tilde{\zeta}(\tilde{t})\tilde{\zeta}(\tilde{t'})\right> = 2\tilde{\xi}\tilde{T}\delta(\tilde{t}-\tilde{t'})$.
The dimensionless underdamped Langevin equations read
\begin{align}
\frac{d\tilde{x}}{d\tilde{t}} &= \tilde{p},\label{eq_dx_non}\\
\frac{d\tilde{p}}{d\tilde{t}} &= -\frac{\partial \tilde{V}}{\partial \tilde{x}} -\tilde{\xi}\tilde{p} + \tilde{\zeta}(\tilde{t}).\label{eq_dp_non}
\end{align}
Also, the dimensionless overdamped Langevin equation reads
\begin{align}
\frac{d\tilde{x}}{d\tilde{t}}= -\frac{1}{\tilde{\xi}}\frac{\partial \tilde{V}}{\partial \tilde{x}} + \frac{1}{\tilde{\xi}}\tilde{\zeta}(\tilde{t}).\label{eq_dx_od_non}
\end{align}
The dimensionless metric tensor is also given by
\begin{align}
    \tilde{g}_{TT}&\equiv m^{-\frac{1}{2}}k_e^{\frac{1}{2n}}k_{\rm B}^{-\frac{n+1}{2n}} T_e^{\frac{3n-1}{2n}} g_{TT},\\
    \tilde{g}_{Tk}&\equiv m^{-\frac{1}{2}}k_e^{\frac{2n+1}{2n}}k_{\rm B}^{-\frac{n+1}{2n}} T_e^{\frac{n-1}{2n}} g_{Tk},\\
    \tilde{g}_{kT}&\equiv m^{-\frac{1}{2}}k_e^{\frac{2n+1}{2n}}k_{\rm B}^{-\frac{n+1}{2n}} T_e^{\frac{n-1}{2n}} g_{kT},\\
    \tilde{g}_{kk}&\equiv m^{-\frac{1}{2}}k_e^{\frac{4n+1}{2n}}k_{\rm B}^{-\frac{n+1}{2n}} T_e^{-\frac{n+1}{2n}} g_{kk}.
\end{align}
With this, the expected form of the metric tensor is
\begin{equation}
    \tilde{g}_{\mu\nu}=\tilde g_{\mu\nu}^p+\tilde g_{\mu\nu}^x = \frac{a_{\mu\nu}}{\tilde{\xi}} + b_{\mu\nu}\tilde{\xi},\label{eq_g_fit}
\end{equation}
if the decomposition in Eqs.~\eqref{eq_dem1}--\eqref{eq_dem3} is feasible.
Meanwhile, for the metric tensor $g_{\mu\nu}^{\rm od}$ of the overdamped dynamics, the expected form should be
\begin{equation}
    \tilde{g}_{\mu\nu}^{\rm od} = c_{\mu\nu}\tilde{\xi},
\end{equation}
in the dimensionless version.

We performed the numerical simulation in the following way.
For a given set of $(\tilde T, \tilde k)$, we first prepared a large number of samples $(\tilde x, \tilde p)$ from the equilibrium distribution by using the accept-reject sampling method~\cite{martino2018accept}.
Then, we simulated each $(\tilde x, \tilde p)$ according to the Langevin equations \eqref{eq_dx_non} and \eqref{eq_dp_non} or Eq.~\eqref{eq_dx_od_non} with long enough simulation time $\tilde{t}_s$ and small enough time step $d\tilde{t}$.
With these $(\tilde x, \tilde p)$, we calculated the fluctuations of thermodynamic forces and the time-correlation functions at each time by taking ensemble averages.
Finally, we obtained $\tilde{g}_{\mu\nu}$ by numerical integration of the time-correlation functions.
We used $4\times 10^5$ samples for the initial equilibrium distribution, and performed the numerical simulations using Euler-Maruyama method with $\tilde{t}_s=100$ and $d\tilde{t}=0.01$.

%\begin{widetext}
\begin{figure*}[t!]
    \includegraphics[width=\linewidth]{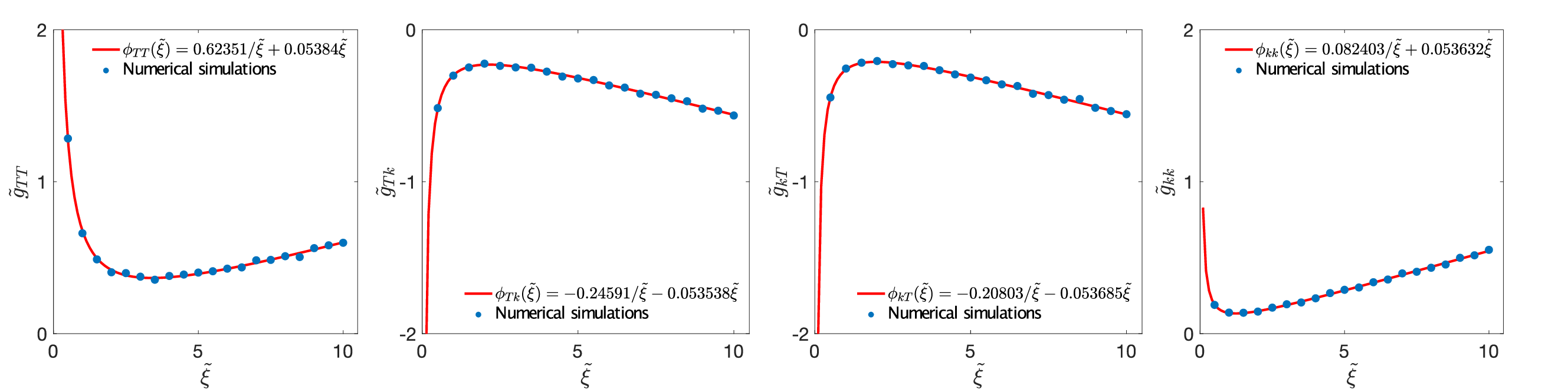}
    \caption{$\tilde \xi$ dependence of the dimensionless metric tensor $\tilde{g}_{\mu\nu}$ at $(\tilde{T},\tilde{k})=(1,1)$ for the scale-invariant potential $V = kx^4/4$. Simulation results (dots) were fitted by the functions $\phi_{\mu\nu} = a_{\mu\nu}/\tilde{\xi} + b_{\mu\nu}\tilde{\xi}$ (solid curves) by using the least square method.}
    \label{fig_g_x4}
\end{figure*}
%\end{widetext}
%\begin{widetext}
\begin{figure*}[t!]
    \includegraphics[width=\linewidth]{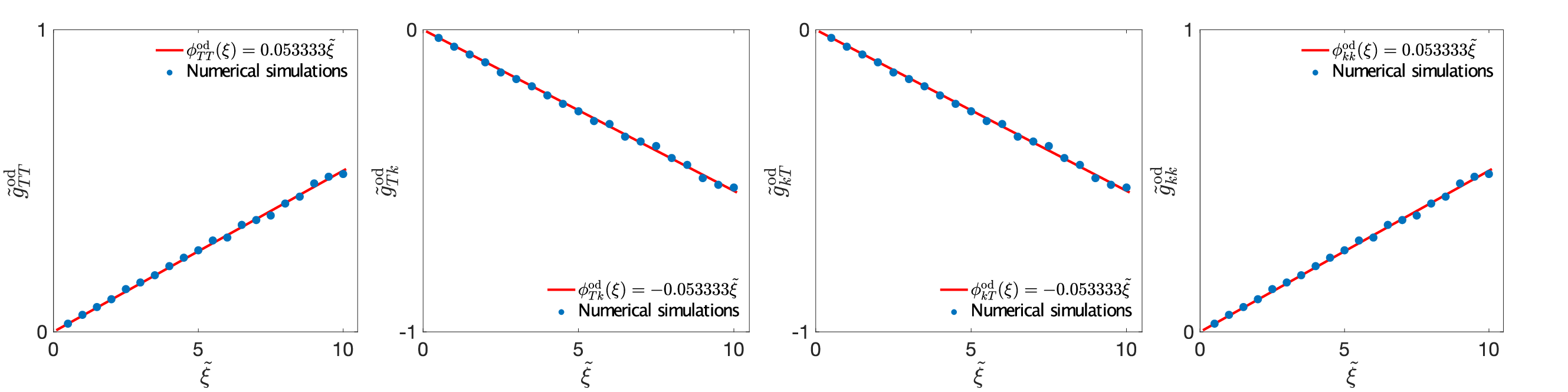}
    \caption{$\tilde \xi$ dependence of the dimensionless overdamped $\tilde{g}_{\mu\nu}^{\rm od}$ at $(\tilde{T},\tilde{k})=(1,1)$ for the scale-invariant potential $V = kx^4/4$. Simulation results (dots) were fitted by the functions $\phi_{\mu\nu}^{\rm od} = c_{\mu\nu}\tilde{\xi}$ (solid curves) by using the least square method.}
    \label{fig_g_od_x4}
\end{figure*}
%\end{widetext}

\autoref{fig_g_x4} 
shows the simulation results of $\tilde{g}_{\mu\nu}$ as a function of $\tilde \xi$ for the scale-invariant potential $V=kx^4/4$ ($n=2$) with $(\tilde{T},\tilde{k})=(1,1)$.
We can find that $\tilde{g}_{\mu\nu}$ is well fitted by $\phi_{\mu\nu} = a_{\mu\nu}/\tilde{\xi} + b_{\mu\nu}\tilde{\xi}$ with $a_{\mu\nu}$ and $b_{\mu\nu}$ being the coefficients,
where $a_{\mu\nu} \simeq a_{\nu\mu}$ and $b_{\mu\nu} \simeq b_{\nu\mu}$~\footnote{There is a slight symmetry breaking for $\phi_{\mu\nu}$, which mainly happens at the small $\tilde{\xi}$ regime. In this regime, the large momentum relaxation timescale~\eqref{eq.tau_p} requires longer simulation time, which is not affordable.}.
Thus, we have confirmed Eq.~(\ref{eq_g_fit}).
We also find the degenerating behavior of $a_{\mu\nu}$, which suggests the degeneracy of $\tilde{g}_{\mu\nu}^p$ as similar to the case of the harmonic potential.

Meanwhile, \autoref{fig_g_od_x4} shows $\tilde{g}_{\mu\nu}^{\rm od}$ calculated by using the overdamped Langevin dynamics Eq.~(\ref{eq_dx_od_non}), 
which is well fitted by $\phi_{\mu\nu}^{\rm od} = c_{\mu\nu}\tilde{\xi}$ with $c_{\mu\nu}$ being the coefficients with $ c_{\mu\nu}\simeq  c_{\nu\mu}$. 
We also find the degenerating behavior of $c_{\mu\nu}$, which was shown in~\cite{watanabe2022finite} for the scale-invariant potentials.
In the sufficiently large $\tilde \xi$ regime, we can also find $b_{\mu\nu} \simeq c_{\mu\nu}$, implying $\tilde g_{\mu\nu}\simeq \tilde{g}_{\mu\nu}^x\simeq \tilde g_{\mu\nu}^{\rm od}$ 
in the overdamped limit as expected (see also Appendix~\ref{appendix_overdamped}).
We also checked that the same behaviors appear for the case of $V=kx^6/6$ ($n=3$) (data not shown).

Together with the timescales identified in Eqs.~\eqref{eq.tau_p} and \eqref{eq_tau_x}, these findings give numerical evidence that the decomposition in Eqs.~\eqref{eq_dem1}--\eqref{eq_dem3} is feasible for the scale-invariant potentials.

\section{Compatibility of Carnot efficiency and finite power}
\label{sec_compatibility}

As an application of the decomposition of metric tensor in \autoref{sec_decomposition}, we consider the compatibility between Carnot efficiency and finite power~\cite{holubec2018cycling,miura2021compatibility,miura2022achieving},
where the efficiency $\eta$ of a heat engine is defined by
\begin{equation}
\eta \equiv \frac{W}{Q_{\rm in}}
\end{equation}
with $Q_{\rm in}$ and $W$ being heat intake and work output, respectively.

Such compatibility may usually be forbidden by the trade-off relations between efficiency and power~\cite{shiraishi2016universial,pietzonka2018universal,dechant2018entropic,brandner2020thermodynamic,hino2021geometrical,vu2024geometric}.
However, it is pointed out that the compatibility can be achieved under the vanishing limit of relaxation timescales~\cite{holubec2018cycling,miura2021compatibility,miura2022achieving}.
The decomposition of metric tensor in \autoref{sec_decomposition} implies that the thermodynamic length, as well as the dissipated availability, may be vanishing under such limit, leading to zero energy loss with finite power.

In particular, in the slow-driving linear response regime \eqref{eq_linear_response}, the trade-off relation between effective efficiency $\varepsilon$ and power $W/t_{\rm cyc}$
for a heat engine operating with cycle time $t_{\rm cyc}$ holds~\cite{brandner2020thermodynamic,hino2021geometrical}. 
Here, the effective efficiency 
\begin{equation}
    \varepsilon\equiv \frac{W}{U}
\end{equation}
characterizes the ratio between the work output of the heat engine
\begin{equation}
    W \equiv \int_0^{t_{\rm cyc}} \left<X_i\right>\dot{\Lambda}^i dt
\end{equation}
and the net heat intake~\cite{brandner2020thermodynamic,frim2022geometric,frim2022optimal,hino2021geometrical}
\begin{equation}
    U \equiv -\int_0^{t_{\rm cyc}} \left<X_T\right>\dot{T} dt.
\end{equation}
By defining the quasi-static work
\begin{equation}
    \mathcal{W} \equiv \oint \mathcal{X}_i d\Lambda^i, \label{eq.qs_work}
\end{equation}
and noting 
\begin{equation}
    \varepsilon \simeq 1-\frac{\mathcal{A}}{\mathcal{W}}\label{eq.varepsilon}
\end{equation}
in the slow-driving linear response regime,
we obtain the following trade-off relation between the effective efficiency $\varepsilon$ and power $\mathcal W/t_{\rm cyc}$~\cite{brandner2020thermodynamic,hino2021geometrical}:
\begin{equation}
    \frac{\mathcal{W}}{t_{\rm cyc}} \leq (1-\varepsilon)\left(\frac{\mathcal{W}}{\mathcal{L}}\right)^2.\label{eq_trade_lin}
\end{equation}
Because of $\eta/\eta_{\rm C}\le \varepsilon$, Eq.~(\ref{eq_trade_lin}) can also be considered the trade-off relation between efficiency $\eta$ and power $\mathcal W/t_{\rm cyc}$~\cite{brandner2020thermodynamic}. 

The trade-off relation~\eqref{eq_trade_lin} forbids the compatibility of Carnot efficiency and finite power in usual cases.
However, cases with $\mathcal{L} \to 0$ are exceptions.
This is easy to understand, as the vanishment of the thermodynamic length implies the vanishment of dissipated availability from Eq.~\eqref{eq_diss}.
By considering the decomposition of the metric tensor in Eqs.~\eqref{eq_dem1}--\eqref{eq_dem3}, it is expected that the vanishing limit of relaxation times realizes such cases.

\begin{figure}
    \includegraphics[width=0.8\linewidth]{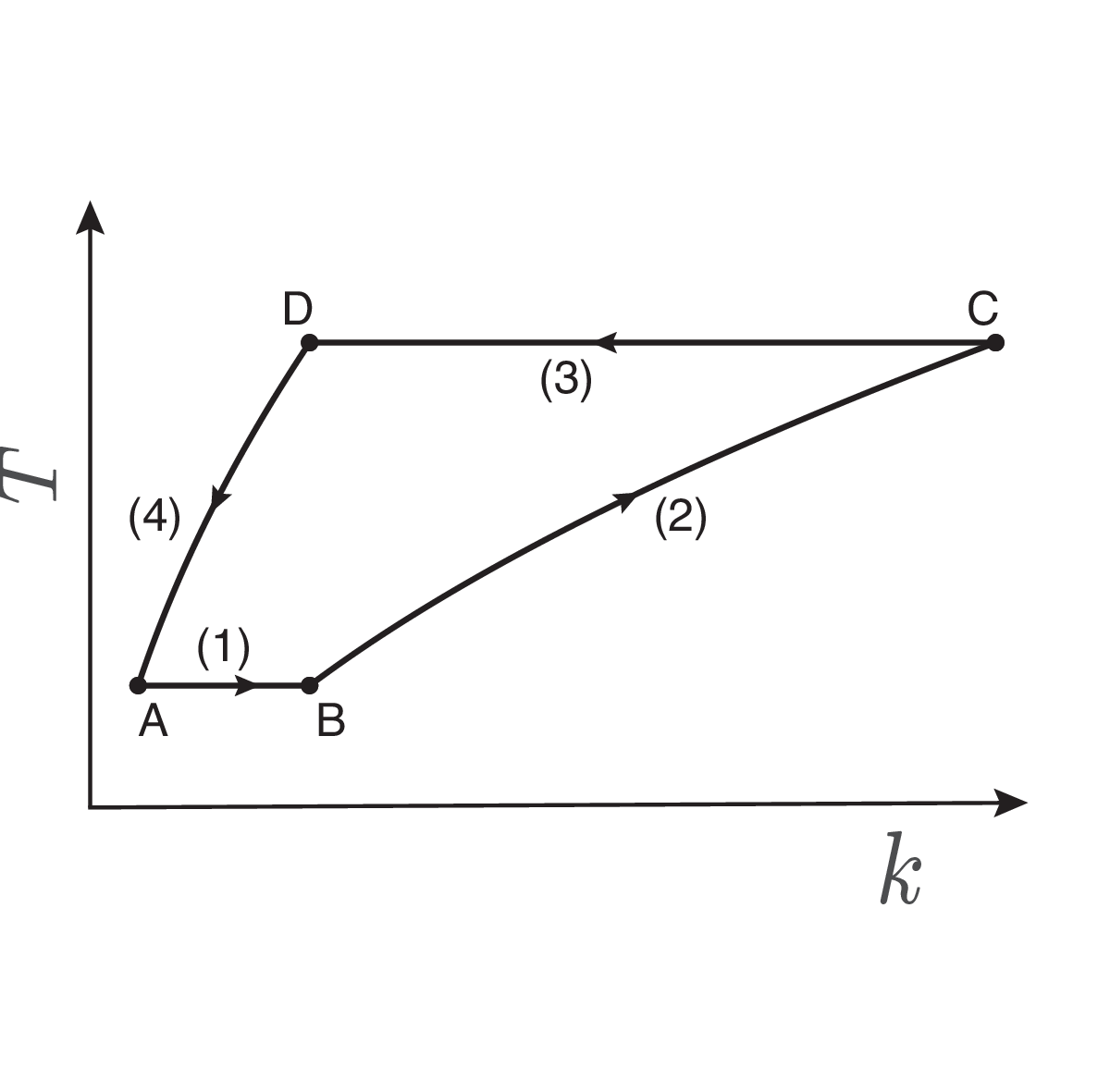}
    \caption{Schematic illustration of the Carnot cycle with (1) isothermal process; (2) adiabatic process; (3) isothermal process; and (4) adiabatic process in $T$--$k$ space, where a Brownian particle in a controllable harmonic potential $V=kx^2/2$ is used as a working substance.}\label{fig_carnot}
\end{figure}

 We demonstrate the above consideration by constructing a Carnot cycle of a Brownian particle in a harmonic potential $V=kx^2/2$, 
 of which the metric tensor is given by~\eqref{eq_g_harmonic}~\cite{frim2022optimal,frim2022geometric,zulkowski2012geometry}.
We construct the Carnot cycle as follows (\autoref{fig_carnot}):
\begin{enumerate}
\item[(1)] Isothermal process from $A(T_0,k_0)$ to $B(T_0,c_1k_0)$:
\begin{equation}
    \begin{aligned}
        T(t)&=T_0,\\
        k(t)&= (c_1 - 1)k_0 t/t_1  + k_0.
    \end{aligned}\label{eq_iso1_process}
\end{equation}
\item[(2)] Adiabatic expansion process from $B(T_0,c_1k_0)$ to $C(c_2 T_0, c_1c_2^2k_0)$:
\begin{equation}
    \begin{aligned}
        T(t) &= (c_2 - 1)T_0 t/t_2  + T_0,\\
        k(t) &= c_1k_0[T(t)/T_0]^2.
    \end{aligned}
\end{equation}
\item[(3)] Isothermal process from $C(c_2 T_0, c_1c_2^2k_0)$ to $D(c_2T_0,c_2^2k_0)$:
\begin{equation}
    \begin{aligned}
        T(t) &= c_2 T_0,\\
        k(t) &= c_2^2(1-c_1)k_0t/t_3 + c_1c_2^2k_0.
    \end{aligned}
\end{equation}
\item[(4)] Adiabatic compression process from $D(c_2T_0,c_2^2k_0)$ to $A(T_0,k_0)$:
\begin{equation}
    \begin{aligned}
        T(t) &= (1-c_2)T_0 t/t_4  + c_2T_0,\\
        k(t) &= k_0[T(t)/T_0]^2.
    \end{aligned}\label{eq_ad2_process}
\end{equation}
\end{enumerate}
Here, $c_1>1$ and $c_2 > 1$ are two constants and $t_i\ (i=1,2,3,4)$ is the duration of each process.
It is obvious that $t_{\rm cyc} = t_1 + t_2 + t_3 + t_4$, $T_c = T_0$, $T_h = c_2T_0$, and the corresponding Carnot efficiency is given by
\begin{equation}
    \eta_{\rm C} = \frac{c_2 -1}{c_2}.\label{eq_carnot_eff}
\end{equation}
It should be noted that the aforementioned condition $c_1>1$ and $c_2>1$ 
imply that the points $A$, $B$, $C$, and $D$ in the $T$--$k$ space in \autoref{fig_carnot} are distinguishable.

Now, we turn to the dissipated availability.
Due to Eq.~\eqref{eq_diss} and the proposed decomposition of the metric tensor~\eqref{eq_dem1}--\eqref{eq_dem3}, it is possible to achieve the vanishment of the dissipated availability with finite operation time by taking the vanishing limit of relaxation times.
In fact, the dissipated availability $\mathcal{A}$ for the cycle in Eqs.~\eqref{eq_iso1_process}--\eqref{eq_ad2_process} is given by
%\begin{widetext}
\begin{align}
    \mathcal{A} &= \sum_{i=1}^4\int_0^{t_i}g_{\mu\nu}\dot{\Lambda}^\mu\dot{\Lambda}^\nu dt = \mathcal{A}_p + \mathcal{A}_x,\label{eq.A_decom}\\
    \mathcal{A}_{p} &\equiv \tau_p \left[\frac{k_{\rm B} T_0 (c_1-1)^2}{4 t_1 c_1}+\frac{k_{\rm B} T_0 (c_1-1)^2c_2}{4 t_3 c_1}\right],\\
    \mathcal{A}_{x} &\equiv \tau_x \left[\frac{ k_{\rm B} T_0 (c_1-1)^2(c_1+1)}{4 t_1 c_1}\right.+\left.\frac{ k_{\rm B} T_0 (c_2 - 1)^2(c_2 + 1)}{8 t_2 c_1 c_2^2}\right.\notag\\
    &\quad\left.+\frac{ k_{\rm B} T_0 (c_1-1)^2(c_1+1)c_2}{4 t_3 c_1^2}\right.+\left.\frac{ k_{\rm B} T_0 (c_2 - 1)^2(c_2 + 1)}{8 t_4 c_2^2}\right],
\end{align}
%\end{widetext}
which is decomposed according to the momentum relaxation time $\tau_p$ and position relaxation time $\tau_x$:
\begin{align}
\tau_p=\frac{m}{\xi}, \ \ \tau_x=\frac{\xi}{k_0}.
\end{align}
The decomposition of the dissipated availability according to the relaxation times in Eq.~(\ref{eq.A_decom}) results from the decomposition in Eqs.~\eqref{eq_dem1}--\eqref{eq_dem3} for the metric tensor.
If we take an appropriate vanishing limit $\tau_p \to 0$ and $\tau_x \to 0$, 
we have $\mathcal{A}\to 0$, thus achieving $\eta \to \eta_{\rm C}$ with finite operation time $t_{\rm cyc}$.

To realize the above scenario, we need to specify the way of taking the simultaneous vanishing relaxation times $\tau_p \to 0$ and $\tau_x \to 0$ as it may significantly affect the underlying dynamics. 
Here, we adopt the following vanishing limit of the relaxation times:
\begin{align}
    \tau_p &= \frac{m}{\xi} = \frac{\theta_p}{\epsilon} \to 0 \ (\epsilon \to +\infty),\label{eq_vanishing_1}
    \\
    \tau_x &= \frac{\xi}{k_0} = \frac{\theta_x}{\epsilon} \to 0 \ (\epsilon \to +\infty),\label{eq_vanishing_2}
\end{align}
respectively, where $\theta_p$ and $\theta_x$ are two positive finite constants, and $\epsilon$ is a dimensionless parameter.
This vanishing limit implies $m \sim \epsilon^{-1}$ and $k_0 \sim \epsilon$ with the ratio being kept finite as $\tau_p/\tau_x = \theta_p/\theta_x$.

The efficiency $\eta$ and power $W/t_{\rm cyc}$ for the cycle~\eqref{eq_iso1_process}--\eqref{eq_ad2_process} under the vanishing limit~\eqref{eq_vanishing_1} and \eqref{eq_vanishing_2} are given by (see Appendix~\ref{appendix_Carnot} for the derivation)
\begin{align}
    \eta &=\eta_{\rm C} - \frac{\gamma}{\epsilon},\label{eq.eta_epsilon}\\
    \frac{W}{t_{\rm cyc}} &=\frac{\mathcal{W}}{t_{\rm cyc}} - \frac{\alpha\gamma}{\epsilon},\label{eq.pow_epsilon}
\end{align}
where the quasi-static work $\mathcal{W}$ in Eq.~(\ref{eq.qs_work}) is given by~\cite{frim2022optimal,frim2022geometric}
\begin{align}
    \mathcal{W} &=-\oint \frac{k_{\rm B} T}{2k}dk= \frac{k_{\rm B} T_0(c_2-1)\ln c_1}{2},\label{eq_quasistatic_work}
    %\frac{\mathcal{W}}{t_{\rm cyc}} &= \frac{k_{\rm B} T_0(c_2-1)\ln c_1}{2t_{\rm cyc}},\label{eq_quasistatic_power}
\end{align}
and $\alpha$ and $\gamma$ are positive finite constants as
\begin{align}
    \alpha &\equiv \frac{k_{\rm B} T_0 c_2\ln c_1}{2t_{\rm cyc}},\label{eq_alpha}\\
    \gamma &\equiv \frac{\theta_x}{2c_2 \ln c_1}\left(1-\frac{1}{c_2^2}\right)\left(\frac{1}{c_1 t_2} + \frac{1}{t_4}\right)\notag \\
    &\quad+ \frac{\theta_x}{4 t_1 c_2 \ln c_1}\left(\frac{c_1 - 1}{c_1}\right)^2\left(c_1+1+2c_1\frac{\theta_p}{\theta_x}\right) \notag\\
    &\quad+ \frac{\theta_x}{4t_3 c_2^2 \ln c_1}\left(\frac{c_1 - 1}{c_1}\right)^2\left(c_1+1+2c_1c_2^2\frac{\theta_p}{\theta_x}\right).\label{eq_gamma}
\end{align}
In the vanishing limit of the relaxation times $\tau_p$ and $\tau_x$ ($\epsilon \to \infty$), we find that the efficiency approaches the Carnot efficiency $\eta_{\rm C}$ (Eq.~\eqref{eq.eta_epsilon}), while the power remains the finite value $\mathcal W/t_{\rm cyc}$ (Eq.~\eqref{eq.pow_epsilon}).
Thus, the compatibility of Carnot efficiency and finite power under the vanishing limit of the relaxation times \eqref{eq_vanishing_1} and \eqref{eq_vanishing_2} has been achieved.

We show that this compatibility is consistent with the trade-off relation~\eqref{eq_trade_lin} by 
estimating the ratio between the left- and right-hand sides of~\eqref{eq_trade_lin} 
in the vanishing limit of the relaxation times.
Combined with the definition of effective efficiency~\eqref{eq.varepsilon}, the trade-off relation~\eqref{eq_trade_lin} becomes
\begin{equation}
    \frac{\mathcal{W}}{t_{\rm cyc}} \leq \frac{\mathcal{A}\mathcal{W}}{\mathcal{L}^2},\label{eq_trade_lin2}
\end{equation}
of which both the left and right-hand sides are constants.
For the protocol we adopted in Eqs.~\eqref{eq_iso1_process}--\eqref{eq_ad2_process}, we choose $c_1=c_2=10$ and $t_1=t_2=t_3=t_4$, and 
$\theta_p=\theta_x=\theta$. 
Then, we can numerically evaluate the thermodynamic length $\mathcal L$ in Eq.~\eqref{eq_length} as
\begin{align}
    \mathcal{L}&=\sum_{i=1}^{4}\int_0^{t_i}\sqrt{g_{\mu\nu}\dot{\Lambda}^\mu\dot{\Lambda}^\nu}dt \approx 5.900\sqrt{k_{\rm B}T_0}\sqrt{\frac{\theta}{\epsilon}}.\label{eq_estimate_L}
\end{align}
The numerical result of the dissipated availability~\eqref{eq_diss} in this protocol is also estimated as
\begin{equation}
    \mathcal{A} \approx \frac{272.201}{t_{\rm cyc}} k_{\rm B}T_0 \frac{\theta}{\epsilon}.\label{eq_estimate_A}
\end{equation}
Thus, the ratio between the left- and right-hand sides of Eq.~(\ref{eq_trade_lin2}) is given by
\begin{equation}
    \frac{\mathcal{A}t_{\rm cyc}}{\mathcal{L}^2}\approx 7.820,
\end{equation}
by using Eqs.~\eqref{eq_estimate_L} and~\eqref{eq_estimate_A}.
Note that this is not such a tight bound, as the protocol we chose for the Carnot cycle is not the geodesic in $T$--$k$ space.

Finally, we support the compatibility of Carnot efficiency and finite power by showing its consistency
with the trade-off relation between efficiency and power that applies to general Markov heat engines beyond the linear response regime~\cite{shiraishi2016universial,shiraishi2023introduction}:
\begin{equation}
    \frac{W}{t_{\rm cyc}} \leq \chi T_c \eta (\eta_{\rm C} - \eta).\label{eq_trade}
\end{equation}
Here, $\chi$ is a positive factor depending on the system.
Again, it may still be possible for the power to remain finite even $\eta \to \eta_{\rm C}$, if $\chi$ diverges with a proper rate at the same time.
In fact, it was pointed out that the vanishing limit of relaxation times of a system can lead to the divergence of $\chi$, realizing the compatibility of Carnot efficiency and finite power~\cite{holubec2018cycling,miura2021compatibility, miura2022achieving}.

The $\chi$ factor in the present case is given by (see Appnedix~\ref{appendix_Carnot} for the derivation)
\begin{equation}
    \chi=\omega \epsilon - \vartheta,\label{eq_chi_harmonic}
\end{equation}
where $\omega$ and $\vartheta$ are positive finite constants:
\begin{align}
    \omega &\equiv \frac{c_2^2 k_{\rm B} (t_2 + 3t_3 + t_4)}{3t_{\rm cyc}\theta_p},\label{eq_omega}\\
    \vartheta &\equiv \frac{c_2^2 k_{\rm B} \ln c_1}{2t_{\rm cyc}}.\label{eq_theta}
\end{align}
It is obvious that $\chi$ diverges in the limit of $\epsilon \to \infty$ (Eqs.~\eqref{eq_vanishing_1} and \eqref{eq_vanishing_2}).
Due to this divergence, even when $\eta$ approaches $\eta_{\rm C}$,  $\chi T_0 \eta(\eta_{\rm C} - \eta)$, which corresponds to the right-hand side of the trade-off relation~\eqref{eq_trade}, remains finite as
\begin{align}
    \chi T_0 \eta(\eta_{\rm C} - \eta) &= \omega\gamma\eta_{\rm C} T_0 - \left(\omega \gamma + \vartheta \eta_{\rm C}\right)\frac{\gamma T_0}{\epsilon} + \frac{\vartheta \gamma^2}{\epsilon^2}\notag\\
    &\to \omega\gamma\eta_{\rm C} T_0,\label{eq_chi_limit}
\end{align}
showing that the compatibility of Carnot efficiency and finite power is possible without breaking the trade-off relation.
It is also worth noting that the right-hand side $\chi T_0 \eta(\eta_{\rm C} - \eta)$ in Eq.~(\ref{eq_trade}) may give a loose bound for the power.
In fact, we find that Eq.~(\ref{eq_chi_limit}) is about 30 times larger than $W/t_{\rm cyc}$ for the same parameters we chose to estimate the ratio between the left- and right-hand sides of the trade-off relation~(\ref{eq_trade_lin2}).

\section{Concluding remarks}
\label{sec_discussion}
In this work, we showed the decomposition of the metric tensor, which is a key element to quantify the dissipated availability in the geometrical framework of thermodynamics~\cite{sivak2012thermodynamic,watanabe2022finite,brandner2020thermodynamic,hino2021geometrical}, in terms of the relaxation times characterizing underdamped Langevin dynamics.
This decomposition was demonstrated by the analytically tractable harmonic potential and the scale-invariant potentials by the numerical simulations.
Moreover, we applied the decomposition of the metric tensor to show the compatibility of Carnot efficiency and finite power~\cite{miura2022achieving,holubec2018cycling,miura2021compatibility}.
We took the Carnot cycle using the Brownian particle in the harmonic potential as an example.
We found that the dissipated availability in one cycle can be decomposed according to the relaxation times due to the decomposition of the metric tensor.
The vanishing limit of the relaxation times resulted in the vanishment of the dissipated availability with finite cycle time and thus the compatibility of Carnot efficiency and finite power.
We also showed that this compatibility is consistent with the trade-off relations between efficiency and power.

There remain some future tasks.
A fully theoretical analysis of the dependence of the metric tensor on relaxation times is necessary to support the decomposition for general potentials.
As the different relaxation timescales are shown to be proportional to different orders of friction coefficient by dimensional analysis, the decomposition is expected to be confirmed if the dependence on friction coefficient is theoretically obtained for the metric tensor.

As we noted in Sec.~\ref{sec_intro}, a tight relationship between thermodynamic geometry and optimal transport geometry exists for overdamped dynamics~\cite{chennakesavalu2023unified,zhong2024beyond}. 
A relationship between the two geometries for underdamped dynamics needs further research.
Related to this, it is noteworthy that a trade-off relation between efficiency and power using Wasserstein distance and dynamical activity has recently been formulated~\cite{vu2024geometric}, where the reciprocal of the dynamical activity gives typical timescale of a system.

Last, though we have theoretically shown that the vanishing limit of the relaxation times leads to the compatibility of Carnot efficiency and finite power, we
expect that actual experiments such as~\cite{martinez2016brownian, dago2024reliability} to verify the compatibility, in which the relaxation times of a system are systematically controlled, will be
conducted in the near future.

\begin{acknowledgements}
This work was supported by JSPS KAKENHI Grant Number 22K03450.
\end{acknowledgements}

\appendix
\section{Metric tensor in overdamped limit}\label{appendix_overdamped}
Here, we formally show $g_{\mu\nu}\to g_{\mu\nu}^{\rm od}$ in the overdamped limit from the definition~\eqref{eq_g_def} of the metric tensor using the equilibrium time-correlation function.
For the sufficiently large $\xi$ limit assuring $\tau_p/\tau_x \ll 1$, the Langevin system~\eqref{eq_dx} and \eqref{eq_dp} naturally becomes the overdamped one (Eq.~\eqref{eq_dx_od}) as the inertia $\dot{p}$ is negligible~\cite{kadanoff2000statistical,risken1996fokker,durang2015overdamped}.
Meanwhile, the metric tensor will approach the overdamped one.
For example, for $\mu=\nu=T$, we have
\begin{widetext}
\begin{align}
    g_{TT} &= \frac{1}{k_{\rm B} T}\int_0^{+\infty}ds \left<\delta X_T(s) \delta X_T(0)\right>_{\rm eq} \notag\\
    &=\frac{1}{k_{\rm B} T}\int_0^{+\infty} \frac{1}{T^2}\left<\left[\frac{p^2(s)}{2m} + V(x(s)) - \frac{\left<p^2\right>_{\rm eq}}{2m} - \left<V\right>_{\rm eq}\right] \left[\frac{p^2(0)}{2m} + V(x(0)) - \frac{\left<p^2\right>_{\rm eq}}{2m} - \left<V\right>_{\rm eq}\right]\right>_{\rm eq}ds\notag\\
    &= \frac{1}{k_{\rm B} T}\int_0^{+\infty} \frac{1}{T^2} \left<\left[\frac{p^2(s)}{2m} - \frac{\left<p^2\right>_{\rm eq}}{2m}\right] \left[\frac{p^2(0)}{2m} + V(x(0)) - \frac{\left<p^2\right>_{\rm eq}}{2m} - \left<V\right>_{\rm eq}\right]\right>_{\rm eq} ds\notag\\
    &\qquad + \frac{1}{k_{\rm B} T}\int_0^{+\infty} \frac{1}{T^2} \left<\left[V(x(s)) - \left<V\right>_{\rm eq}\right] \left[\frac{p^2(0)}{2m} + V(x(0)) - \frac{\left<p^2\right>_{\rm eq}}{2m} - \left<V\right>_{\rm eq}\right]\right>_{\rm eq}ds\notag\\
    &\simeq \frac{1}{k_{\rm B} T}\int_0^{+\infty} \frac{1}{T^2} \left<\left[\frac{p^2(0)}{2m} - \frac{\left<p^2\right>_{\rm eq}}{2m}\right] \left[\frac{p^2(0)}{2m} + V(x(0)) - \frac{\left<p^2\right>_{\rm eq}}{2m} - \left<V\right>_{\rm eq}\right]\right>_{\rm eq} \exp\left(-\frac{2\xi}{m}s\right)ds\notag\\
    &\qquad + \frac{1}{k_{\rm B} T}\int_0^{+\infty} \frac{1}{T^2} \left<\left[V(x(s)) - \left<V\right>_{\rm eq}\right] \left[V(x(0)) - \left<V\right>_{\rm eq}\right]\right>_{\rm eq}ds\notag\\
    &= \frac{m}{2\xi}\frac{1}{T^2} \left<\left[\frac{p^2(0)}{2m} - \frac{\left<p^2\right>_{\rm eq}}{2m}\right] \left[\frac{p^2(0)}{2m} + V(x(0)) - \frac{\left<p^2\right>_{\rm eq}}{2m} - \left<V\right>_{\rm eq}\right]\right>_{\rm eq}\notag\\
    &\qquad + \frac{1}{k_{\rm B} T}\int_0^{+\infty} \frac{1}{T^2} \left<\left[V(x(s)) - \left<V\right>_{\rm eq}\right] \left[V(x(0)) - \left<V\right>_{\rm eq}\right]\right>_{\rm eq}ds\notag\\
    &\simeq \frac{1}{k_{\rm B} T}\int_0^{+\infty} \frac{1}{T^2} \left<\left[V(x(s)) - \left<V\right>_{\rm eq}\right] \left[V(x(0)) - \left<V\right>_{\rm eq}\right]\right>_{\rm eq}ds\notag\\
    &= g_{TT}^{\rm od},\label{eq_large_xi}
\end{align}
\end{widetext}
where we used Eq.~\eqref{eq_XT_re} in the second line.
In the above approximation,
rapid damping of $p^2(s)$ in the overdamped dynamics is the key~\cite{kadanoff2000statistical,risken1996fokker,durang2015overdamped}.
This damping can be characterized by
\begin{equation}
    \frac{p^2(s)}{2m} - \frac{\left<p^2\right>_{\rm eq}}{2m} \simeq \left(\frac{p^2(0)}{2m} - \frac{\left<p^2\right>_{\rm eq}}{2m}\right)\exp\left(-\frac{2\xi}{m}s\right),\label{eq_p2_damping}
\end{equation}
which is applied in the fifth line of Eq.~\eqref{eq_large_xi}.
To derive Eq.~\eqref{eq_p2_damping}, we start from Eq.~\eqref{eq_phi(s)}.
By substituting $\phi(s)$ as $(p^2(s)-p^2(0))/(2m)$ and taking the partial derivative of $s$ in both sides, we have
\begin{align}
    \frac{\partial}{\partial s}\left[\frac{p^2(s)}{2m} - \frac{\left<p^2\right>_{\rm eq}}{2m}\right] = \mathcal{L}_{\rm FP}^\dagger \left[\frac{p^2(s)}{2m} - \frac{\left<p^2\right>_{\rm eq}}{2m}\right].\label{eq_p2_dt}
\end{align}
We can decompose $\mathcal{L}_{\rm FP}^\dagger$
\begin{equation}
    \mathcal{L}_{\rm FP}^\dagger = \mathcal{L}_{\rm re}^\dagger + \mathcal{L}_{\rm irr}^\dagger,
\end{equation}
in terms of a reversible part $\mathcal{L}_{\rm re}^\dagger$
\begin{equation}
    \mathcal{L}_{\rm re}^\dagger \equiv \frac{p}{m}\frac{\partial}{\partial x} - \frac{\partial V}{\partial x}\frac{\partial}{\partial p},
\end{equation}
and an irreversible part
\begin{equation}
    \mathcal{L}_{\rm irr}^\dagger \equiv - \frac{\xi}{m}p\frac{\partial}{\partial p} + \xi k_{\rm B} T \frac{\partial^2}{\partial p^2}.
\end{equation}
In the sufficiently large $\xi$ regime, we have
\begin{equation}
    \mathcal{L}_{\rm FP}^\dagger \simeq \mathcal{L}_{\rm irr}^\dagger.
\end{equation}
Combined with
\begin{equation}
    \mathcal{L}_{\rm irr}^\dagger \left[\frac{p^2(s)}{2m} - \frac{\left<p^2\right>_{\rm eq}}{2m}\right]
    = -\frac{2\xi}{m}\left[\frac{p^2(s)}{2m} - \frac{\left<p^2\right>_{\rm eq}}{2m}\right],
\end{equation}
where $\left<p^2\right>_{\rm eq} = mk_{\rm B}T$, Eq.~\eqref{eq_p2_dt} becomes
\begin{align}
    \frac{\partial}{\partial s}\left[\frac{p^2(s)}{2m} - \frac{\left<p^2\right>_{\rm eq}}{2m}\right] \simeq -\frac{2\xi}{m}\left[\frac{p^2(s)}{2m} - \frac{\left<p^2\right>_{\rm eq}}{2m}\right].\label{eq_p2_dt_od}
\end{align}
Equation~\eqref{eq_p2_damping} can be obtained by solving Eq.~\eqref{eq_p2_dt_od}.

Moreover, we also apply
\begin{equation}
    \left<\left[V(x(s)) - \left<V\right>_{\rm eq}\right] \left[\frac{p^2(0)}{2m} - \frac{\left<p^2\right>_{\rm eq}}{2m} \right]\right>_{\rm eq}=0,
\end{equation}
in the sixth line of Eq.~\eqref{eq_large_xi}, as $V(x(s))$ does not depend on $p(0)$ in overdamped dynamics.

Such analysis can also be applied for other components, which shows that the metric tensor approaches the overdamped one in the sufficiently large $\xi$ limit.

\section{Proportionality of relaxation time in metric tensor for overdamped dynamics}\label{appendix_overdamped_tau}
We show $g_{\mu\nu}^{\rm od}\propto \tau_x$ for general potentials. To this end, we consider its dependence on $\xi$ as
\begin{align}
    \frac{\partial g_{\mu\nu}^{\rm od}}{\partial \xi} &= \frac{1}{k_{\rm B}T}\int_0^{+\infty}ds\frac{\partial}{\partial \xi}\left<\delta X_\mu(s) \delta X_\nu (0)\right>_{\rm eq}\notag\\
    &= \frac{1}{k_{\rm B}T}\int_0^{+\infty}ds\left<\left[\frac{\partial}{\partial \xi}e^{\mathcal{G}_{\rm FP}^\dagger s} \delta X_\mu (0)\right] \delta X_\nu (0)\right>_{\rm eq}\notag\\
    &= -\frac{1}{k_{\rm B}T \xi}\int_0^{+\infty}ds\left<\left[s\mathcal{G}_{\rm FP}^\dagger e^{\mathcal{G}_{\rm FP}^\dagger s} \delta X_\mu (0)\right] \delta X_\nu (0)\right>_{\rm eq}\notag\\
    &= -\frac{1}{k_{\rm B}T \xi}\int_0^{+\infty} s\frac{d}{ds}\left<\delta X_\mu(s) \delta X_\nu (0)\right>_{\rm eq}ds\notag\\
    &= \frac{1}{k_{\rm B}T \xi}\int_0^{+\infty}ds\left<\delta X_\mu(s) \delta X_\nu (0)\right>_{\rm eq}\notag\\
    &= \frac{g_{\mu\nu}^{\rm od}}{\xi},\label{eq_od_xi} 
\end{align}
where we applied Eq.~\eqref{eq_phi(s)} and the adjoint Fokker-Planck operator $\mathcal{G}_{\rm FP}^\dagger$ for the overdamped dynamics in Eq.~\eqref{eq_ad_fp_od} to obtain $\delta X_{\mu}(s)$.
Following derivatives,
\begin{align}
    \frac{\partial}{\partial \xi}e^{\mathcal{G}_{\rm FP}^\dagger s} &= -\frac{1}{\xi}s\mathcal{G}_{\rm FP}^\dagger e^{\mathcal{G}_{\rm FP}^\dagger s},\\
    \frac{d}{ds}e^{\mathcal{G}_{\rm FP}^\dagger s} &= \mathcal{G}_{\rm FP}^\dagger e^{\mathcal{G}_{\rm FP}^\dagger s},
\end{align}
are also applied in the third and fourth lines, respectively.
Integration by parts and the following relation
\begin{equation}
    \lim_{s\to +\infty} s\left<\delta X_\mu(s) \delta X_\nu (0)\right>_{\rm eq} = 0,
\end{equation}
which ensures the convergence of the integration in Eq.~\eqref{eq_g_def}, are used in the fifth line.
By solving Eq.~\eqref{eq_od_xi}, we obtain $g_{\mu\nu}^{\rm od}=K_{\mu\nu}\xi$ with $K_{\mu\nu}$ being a matrix independent of $\xi$.

If the potential $V$ can be described in a series form,
\begin{equation}
    V = \sum_{n=1}^{\infty}\frac{k_n}{n}x^{n},
\end{equation}
we can obtain a set of timescales $\tau_n$ without mass $m$ for all $k_n \neq 0$ by dimensional analysis:
\begin{eqnarray}
    \tau_n \sim \xi k_n^{-\frac{2}{n}}(k_{\rm B} T)^{-\frac{n-2}{n}}.
\end{eqnarray}
All these timescales and their combinations $\tau_i^{\alpha}\tau_j^{1-\alpha}$ are proportional to $\xi$, which implies that the position relaxation timescale is proportional to $\xi$:
\begin{equation}
    \tau_x \propto \xi.
\end{equation}
Thus, the metric tensor for overdamped dynamics should be proportional to $\tau_x$ as $g_{\mu\nu}^{\rm od}$ is proportional to $\xi$.

\section{Derivation of Eq.~(\ref{eq.eta_epsilon}), Eq.~(\ref{eq.pow_epsilon}), and Eq.~(\ref{eq_chi_harmonic})}\label{appendix_Carnot}
We show the derivation of the efficiency $\eta$ in Eq.~(\ref{eq.eta_epsilon}), power $W/t_{\rm cyc}$ in Eq.~(\ref{eq.pow_epsilon}), and $\chi$ factor in Eq.~\eqref{eq_chi_harmonic} for the Carnot cycle~\eqref{eq_iso1_process}--\eqref{eq_ad2_process}.

The linear response relation~\eqref{eq_linear_response} is applicable when the driving speed is much slower than the system's relaxation speed, which can be applied in the vanishing limit of relaxation times.
The linear response coefficient, which also serves as the metric tensor, is given by Eq.~\eqref{eq_g_harmonic} if we choose the control variables as $\Lambda^\mu = (T,k)$.
Thus, for $\left<X_k\right>$, we have
\begin{align}
    \left<X_k\right> &= \mathcal{X}_k - g_{k\mu}\dot{\Lambda}^\mu\notag\\
    &= -\frac{k_{\rm B} T}{2k} + \frac{(\xi^2+2mk)k_{\rm B}}{4\xi k^2}\dot{T} - \frac{(\xi^2+mk)k_{\rm B} T}{4\xi k^3}\dot{k},\label{eq_fluc_Xk}
\end{align}
and the work of the Carnot cycle~\eqref{eq_iso1_process}--\eqref{eq_ad2_process} reads
\begin{align}
    W &= \sum_{i=1}^4\int_0^{t_i} \left<X_k\right>\dot{k}dt \notag\\
    &=\frac{k_{\rm B} T_0 (c_2 - 1)\ln c_1}{2} - \frac{k_{\rm B} T_0\xi}{4k_0}\left(1-\frac{1}{c_2^2}\right)\left(\frac{1}{c_1 t_2} + \frac{1}{t_4}\right)\notag\\
    &\quad -\frac{k_{\rm B} T_0 \xi}{8k_0 t_1}\left(\frac{c_1 - 1}{c_1}\right)^2\left(c_1+1+\frac{2k_0 m c_1}{\xi^2}\right) \notag\\
    &\quad-\frac{k_{\rm B} T_0 \xi}{8k_0 c_2t_3}\left(\frac{c_1 - 1}{c_1}\right)^2\left(c_1+1+\frac{2k_0 m c_1c_2^2}{\xi^2}\right).\label{eq_W}
\end{align}
\par
The heat intake $Q_{\rm in}$ of the Carnot cycle~\eqref{eq_iso1_process}--\eqref{eq_ad2_process} is given by~\cite{miura2022achieving}
\begin{align}
    Q_{\rm in} &= \sum_{Q_i > 0}Q_i,\\
    Q_i &= \int_0^{t_i} \frac{\xi}{m} \left(k_{\rm B} T-\frac{\left<p^2\right>}{m}\right)dt\quad (i=1,2,3,4),
\end{align}
where $Q_2$ and $Q_4$ during the adiabatic processes should vanish.
With the definition of $X_k$ in Eq.~\eqref{eq_Xi_def} and the replacement of $\delta X_T$ in Eq.~\eqref{eq_XT_re}, we have
\begin{equation}
    \left<p^2\right> - \left<p^2\right>_{\rm eq}=2m\left(T\left<\delta X_T\right> + k\left<\delta X_k\right>\right).
\end{equation}
Combined with the linear response relations~\eqref{eq_linear_response}, $\left<p^2\right>$ can be expressed as
\begin{align}
    \left<p^2\right>=mk_{\rm B} T -\frac{m^2 k_{\rm B}}{\xi}\dot{T} +\frac{m^2 k_{\rm B} T}{2\xi k}\dot{k},\label{eq_p_square}
\end{align}
where $\left<p^2\right>_{\rm eq} = mk_{\rm B}T$.
Finally, the heat intake is given by
\begin{equation}
    Q_{\rm in}=Q_3=\frac{k_{\rm B} T_0 c_2\ln c_1}{2}.\label{eq_Qin}
\end{equation}

With the work $W$ in Eq.~\eqref{eq_W} and the heat intake $Q_{\rm in}$ in Eq.~\eqref{eq_Qin}, it is easy to obtain the efficiency $\eta$:
\begin{align}
    \eta &= \frac{W}{Q_{\rm in}}\notag\\
    &=\eta_{\rm C} - \frac{\xi}{2k_0 c_2 \ln c_1}\left(1-\frac{1}{c_2^2}\right)\left(\frac{1}{c_1 t_2} + \frac{1}{t_4}\right)\notag\\
    &\qquad - \frac{\xi}{4k_0 t_1 c_2 \ln c_1}\left(\frac{c_1 - 1}{c_1}\right)^2\left(c_1+1+\frac{2k_0 m c_1}{\xi^2}\right)\notag\\
    &\qquad -\frac{\xi}{4k_0 t_3 c_2^2 \ln c_1}\left(\frac{c_1 - 1}{c_1}\right)^2\left(c_1+1+\frac{2k_0 m c_1c_2^2}{\xi^2}\right).\label{eq_eta_calc}
\end{align}
We can express the efficiency $\eta$ in Eq.~\eqref{eq_eta_calc} and the power $W/t_{\rm cyc}$ with $W$ given in Eq.~\eqref{eq_W} 
in terms of $\epsilon$ as
\begin{align}
    \eta &=\eta_{\rm C} - \frac{\gamma}{\epsilon},\\
    \frac{W}{t_{\rm cyc}} &=\frac{\mathcal{W}}{t_{\rm cyc}} - \frac{\alpha \gamma}{\epsilon},
\end{align}
where $\alpha$ and $\gamma$ are given in Eqs.~(\ref{eq_alpha}) and (\ref{eq_gamma}), respectively.

The $\chi$ factor in Eq.~\eqref{eq_trade} for the case of harmonic potential is given as~\cite{dechant2018entropic,miura2022achieving}
\begin{align}
    \chi &= \frac{\xi T_h^2}{t_{\rm cyc} T_c^2 m^2}\sum_{i=1}^4\int_0^{t_i} \frac{1}{T}\left(\frac{T-T_c}{T_h-T_c}\right)^2 \left<p^2\right>dt\notag\\
    &= \frac{\xi c_2^2}{t_{\rm cyc} m^2}\sum_{i=1}^4\int_0^{t_i} \frac{1}{T}\left(\frac{T-T_0}{c_2T_0-T_0}\right)^2 \left<p^2\right>dt,\label{eq_def_chi}
\end{align}
where the temperatures of heat reservoirs $T_c = T_0$ and $T_h = c_2 T_0$ are set according to the Carnot cycle~\eqref{eq_iso1_process}--\eqref{eq_ad2_process}.
By applying Eq.~\eqref{eq_p_square}, we can further calculate $\chi$ in Eq.~\eqref{eq_def_chi} as
\begin{align}
    \chi=\omega \epsilon - \vartheta,
\end{align}
where $\omega$ and $\vartheta$ are given in Eqs.~(\ref{eq_omega}) and (\ref{eq_theta}), respectively.
\bibliographystyle{apsrev4-2.bst}
%\bibliography{metronome}
%apsrev4-2.bst 2019-01-14 (MD) hand-edited version of apsrev4-1.bst
%Control: key (0)
%Control: author (72) initials jnrlst
%Control: editor formatted (1) identically to author
%Control: production of article title (-1) disabled
%Control: page (0) single
%Control: year (1) truncated
%Control: production of eprint (0) enabled
%

\end{document}